\documentclass[12pt]{iopart}
\makeatletter
\@namedef{ver@amsmath.sty}{}
\makeatother
\usepackage{amstext}
\usepackage[latin1]{inputenc}
\usepackage{amsmath}
\usepackage{amsfonts}
\usepackage{amssymb}
\usepackage{graphicx}

\begin{document}

\title[Dual species oven characterised by spatially resolved fluorescence]{Collimated dual species oven source and its characterisation via spatially resolved fluorescence spectroscopy}

\author{N Cooper, E Da Ros, J Nute, D Baldolini, P Jouve, \& L Hackerm\"{u}ller}
\address{School of Physics and Astronomy, University of Nottingham, University Park, Nottingham, NG7 2RD, UK}

\author{M Langer}
\address{Department of Mathematics and Statistics, University of Strathclyde, 26 Richmond Street, Glasgow G1 1XH, UK}

\ead{nathan.cooper@nottingham.ac.uk}
\vspace{10pt}
\begin{indented}
\item[]December 2017
\end{indented}

\begin{abstract}
We describe the design, construction and characterisation of a collimated, dual-species oven source for generating intense beams of lithium and caesium in UHV environments. Our design produces full beam overlap for the two species. Using an aligned microtube array the FWHM of the output beam is restricted to $\sim75$ milliradians, with an estimated axial brightness of 
3.6$\times 10^{14}$ atoms s$^{-1}$ sr$^{-1}$ for Li and 7.4$\times 10^{15}$ atoms s$^{-1}$ sr$^{-1}$ for Cs. We measure the properties of the output beam using a spatially-resolved fluorescence technique, which allows for the extraction of additional information not accessible without spatial resolution.
\end{abstract}


\section{Introduction}

In-vacuum sources of atomic and molecular beams are essential for matterwave interferometry experiments, in particular in the context of quantum technologies \cite{matterwave1,matterwave2}, for experiments working directly with particle beams \cite{beam1,beam2} and for use with Zeeman slowers and/or magneto-optical traps in atomic physics experiments \cite{stan05,zeeman1}. In this paper we describe the design and construction of an in-vacuum source for the simultaneous production of beams of two alkali metal species, in our case lithium and caesium. We then measure the key properties of this source and compare them with theoretical predictions. In doing so we employ spatially resolved fluorescence measurements, which provide access to a wide range of beam parameters. 

A high degree of collimation is in general desirable in a beam source, as this allows the required axial beam brightness to be achieved with a lower total emission rate. This improves the lifetime of the source, helps to maintain a lower background pressure in the vacuum system during source operation, and reduces the likelihood of chemically aggressive species damaging other components of the vacuum system --- for example caesium exposure can degrade ion pumps and high vacuum pressure gauges, while lithium will corrode copper gaskets \cite{dbl_array}.

Our source employs an aligned microtube array to improve output beam collimation \cite{microtube1,microtube2}. Although alternative methods for beam collimation have been demonstrated --- see \cite{candlestick} for example --- a microtube array offers a good balance between simplicity and effectiveness. Our device represents one of the first cases in which separate atomic species have been mixed \emph{prior} to collimation, as previous dual-species sources have typically either not employed a microtube-based collimation device or have used separate collimation arrays for the two species \cite{stan05, zeeman1, dbl_array, fermitrap}. 
We analyse both the transverse and longitudinal velocity distributions within the atomic beam, using only transverse illumination and observation. We determine the degree of collimation of the output beam and find good agreement with theoretical predictions. In both cases we find that a modification of the mean free path of atoms in the microtube array (to account for the effects of interspecies collisions) is the only alteration needed to accurately describe dual-species emission using the existing single-species theory.

\section{Design and theoretical output properties}

\label{theosec}

\begin{figure}
\begin{center}
 \includegraphics[width = 12 cm]{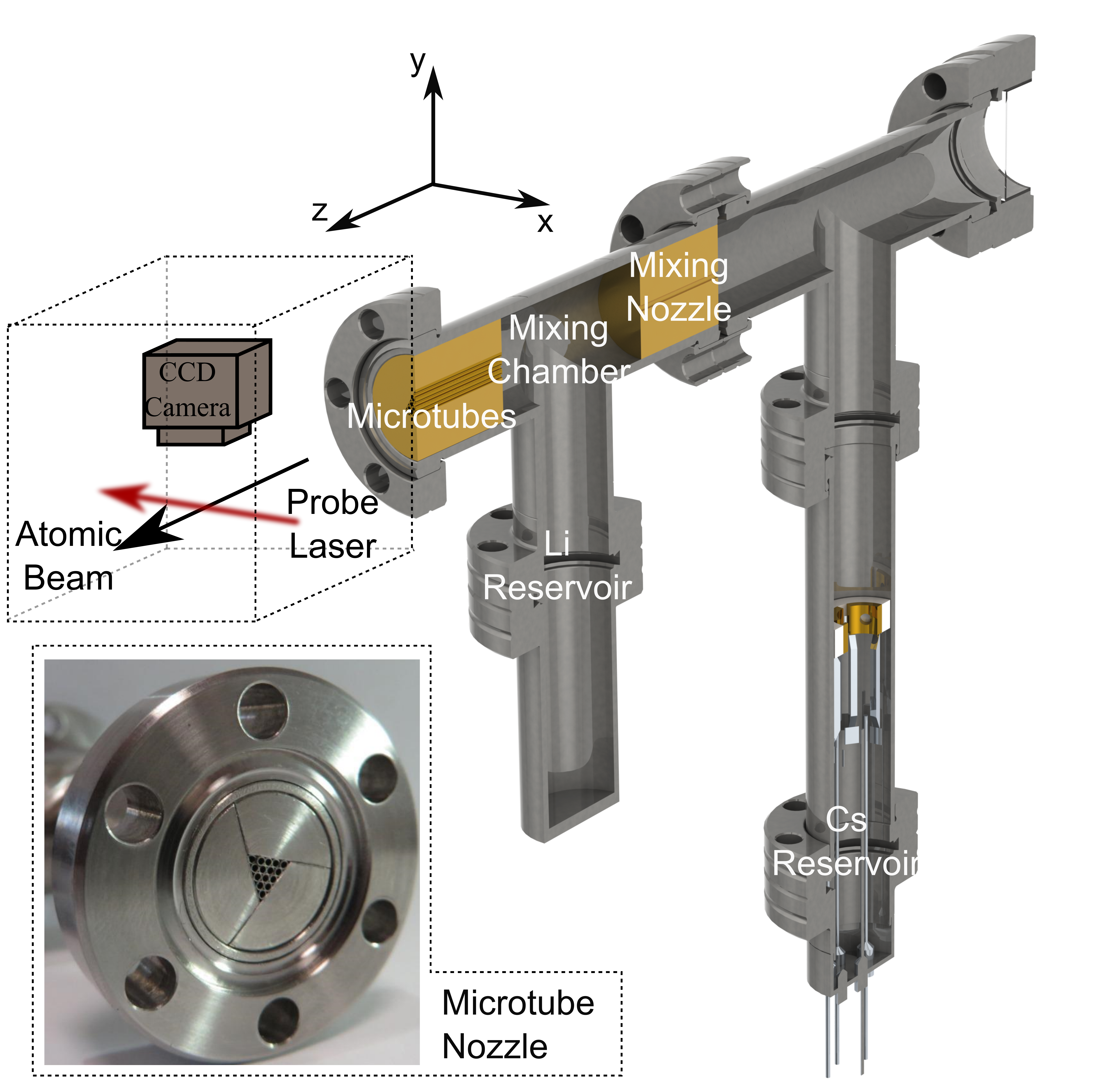}
 \caption{Diagram showing the structure of the dual-species beam source and the setup used for spatially resolved fluorescence spectroscopy of its output. Fluorescence images were captured using the camera shown for a range of different laser detunings, thus enabling reconstruction of the space-velocity distribution of the atoms in the beam. Separate, independently controlled heating wires are used to regulate the temperatures of the Cs and Li reservoirs, the mixing chamber and the microtube array.}
 \label{oven_layout}
 \end{center}
 
\end{figure}

The basic structure of the beam source is illustrated in figure \ref{oven_layout}. Separate, feedback-controlled heating wires allow independent control of the temperatures of the various chambers. Li atoms are sourced by heating a chunk of metallic lithium in the Li reservoir, which directly adjoins the mixing chamber. In the Cs reservoir electrically activated Cs dispensers (SAES getters) are used to provide an initial coating of caesium on the walls of the chamber, following which heating the walls alone can sustain Cs output for several days with the dispensers switched off. The Cs atoms reach the mixing chamber via the mixing nozzle, which is present to prevent backflow of Li atoms into the Cs reservoir. Backflow is prevented by ensuring that sufficient pressure is maintained in the Cs reservoir to cause forward flow through the mixing nozzle and that the mean free path of Li atoms in the mixing nozzle is much shorter than the length of the nozzle, as has been demonstrated in previous work with dual species sources \cite{stan05}. To avoid clogging, the mixing nozzle and the microtube array are maintained at a higher temperature than the Li and Cs reservoirs.    

We use 304-grade stainless steel microtubes of $L=20$ mm length with inner diameter $d_i=0.51$ mm and outer diameter $d_o=0.81$ mm, in a 15-tube triangular array. These are mounted in a custom-built holder which ensures alignment of the tubes and is push-fitted into the CF16 vacuum tube. The viewports providing optical access for the probe laser were heated to $\sim$ 150 $^{\circ}$C during operation to prevent condensation of Li and Cs and thus avoid the formation of a metallic coating on the viewports. The viewport providing access for the camera was heated to $\sim$ 100 $^{\circ}$C, to avoid thermal damage to the camera. 

A theoretical model for emission of particles from an extended tube for a single species was developed primarily by Clausing \cite{clausing}, and is now a well-known result \cite{scoles, atmol_beams}. Firstly, the rate of particles entering the tube per unit area (at the reservoir end with temperature T) is denoted by $\nu_0 = n_0 \bar{v}/4$. The density $n_0$ is given by the vapour pressure and the mean thermal speed of the particles, $\bar{v}$, is equal to $\sqrt{2k_{B}T/m}$, with $k_B$ the Boltzmann constant and $m$ the mass of the particles. An angular term $j_{\theta}(\theta)$\footnote[1]{Throughout this article we denote normalised density functions by lower case letters and absolute density functions by capitals. The variables over which we consider the density distribution are denoted by subscript letters, and any relevant variables not present are assumed to be integrated over such that $F_{a}$ is equal to the integral of $F_{a,b}$ over $b$ etc.} is then defined such that the two-dimensional flux density, $I_{\theta,\phi}(\theta,\phi)$ (from now on shortened to $I_{\theta}(\theta)$ as circular symmetry removes the $\phi$ dependence), is given by
\begin{equation}
 I_{\theta}(\theta) = \frac{\nu_0 A j_{\theta}(\theta)}{\pi},
 \label{Itheta}
\end{equation}
with $A$ being the cross-sectional area of the tube. The form of $j_{\theta}(\theta)$ is dependent on the Knudsen number, $K_{nL}=\lambda_P/L$. $L$ is the length of the tube. The mean free path length for species $a$, $\lambda_{P,a}$, is given by $\lambda_{P,a}^{-1}=\sum_i \sqrt{2}\pi n_i \sigma_{a,i}^{2}$, with $n_i$ the density of particles of species $i$ and $\sigma_{a,i}$ the relevant interparticle collision cross section. This relation can be obtained by adapting the standard single-species expression \cite{microtube1} such that the probability of collision per unit distance travelled is the sum of that arising from all of the different particle species present, thus making the inverse of the overall mean free path equal to the sum of the inverses of the mean free paths against collision with each of the relevant particle species. We calculate $\sigma_{a,i}$ using the geometrical collision cross section $\sigma_{a,i} = \pi (r_a+r_i)^2$, where $r_a$ and $r_i$ are the mean atomic radii of the particles $a$ and $i$. The mean atomic radii for lithium and caesium are 152 pm and 265 pm respectively.

The properties of an atomic beam source are strongly dependent on the Knudsen number, and by convention the range of possible Knudsen numbers is divided into three regimes: transparent, opaque and viscous flow. The two relevant to most atomic beam sources are the `transparent' regime where $K_{nL} > 10$ and the `opaque' regime, defined such that $\beta = 2r/L < K_{nL} < 10$, with $r$ the radius of the tube. Based on the calculated mean free paths above, we expect our source to operate in the opaque regime for both Li and Cs.
For sources operating in the transparent or opaque regimes, while $\tan{\theta} \leq \beta$, the relative flux density $j_{\theta}(\theta)$ can be approximated by
\begin{equation}
 \fl  j_{\theta}(\theta) = \xi_{0} \cos{\theta} + \frac{2 e^{\delta'^{2}}}{\delta' \sqrt{\pi}}\xi_{0}\cdot \cos{\theta} \left( \frac{R(q)}{2}\left[\text{erf}\left(\frac{\delta'\xi_1}{\xi_0}\right)-\text{erf}(\delta')\right] + F(\xi_0,\xi_1,\delta') + S(q)\right),
 \label{jthetalong}
\end{equation}
where $q = \beta^{-1} \tan{\theta}$ and $R(q) = \arccos{q} - q \sqrt{1-q^2}$.
The parameters $\xi_i$ are related to the assumed form of the variation of particle number density with position along the tube, and they can be approximated by $\xi_0 = \alpha$ and $\xi_1 = 1-\alpha$ \cite{olander}, where
\begin{equation}
 \alpha = \frac{1}{2} - \frac{1}{3 \beta^2} \left( \frac{1-2 \beta^3 +(2 \beta^2 -1)\sqrt{1+\beta^2}}{\sqrt{1+\beta^2}-\beta^2 \sinh{^{-1}(1/\beta)}} \right).
\end{equation}  
The remaining undefined terms in eq. (\ref{jthetalong}) are $\delta = \frac{\xi_0}{\sqrt{2 K_{nL}(\xi_1 - \xi_0)}}$ and $\delta' = \frac{\delta}{\sqrt{\cos{\theta}}}$ as well as 
\begin{equation}
 F(\xi_0,\xi_1,\delta') = \frac{2(1-\xi_1)}{\xi_0 \sqrt{\pi}} \delta' \text{exp}(-\delta'^2\xi_1^2/\xi_0^2)
\end{equation}
and
\begin{equation}
 S(q) = \int_0^q \sqrt{1-t^2} \left[ \text{erf}\left(\delta'+\frac{\delta'  t (\xi_1-\xi_0)}{q \xi_0}\right) - \text{erf}(\delta') \right].
\end{equation}
For angles larger than the maximal angle given by the aspect ratio, such that $\tan{\theta} \geq \beta$, $j_{\theta}(\theta)$ is described by 
\begin{equation}
 j_{\theta}(\theta) = \xi_0 \cos{\theta} + \frac{2 e^{\delta'^{2}}}{\delta' \sqrt{\pi}} \xi_{0} S(1) \cos{\theta}.
 \label{endtheo}
\end{equation}
These equations allow us to make a theoretical prediction for the angular distribution of particles emitted from a single tube source, and are valid in both the transparent and opaque regimes. 

The degree of collimation produced by a source is typically characterised by a `peaking factor' $\kappa$ \cite{scoles}, defined as $\pi$ times the ratio of the axial beam brightness $I_{\theta}(0)$(atoms per second per steradian) to the total emission rate $I_{\mathrm{total}}$ (atoms per second). An expression for the value of $\kappa$ in the limit $\beta \ll 1$ was derived in \cite{clausing}:
\begin{equation}
 \kappa = \frac{\pi I_{\theta}(0)}{I_{\mathrm{total}}}= \frac{1}{2}\left( 1+\frac{\pi^{1/2}e^{\delta^2}(1-\text{erf}(\delta))}{2 \delta} \right).
 \label{kappa}
\end{equation}
We use this expression to derive the expected values of $\kappa$ for Li and Cs emitted from our oven, which turn out to be 29 and 24 respectively. In the transparent limit and when $r \ll L$ the peaking factor simplifies to $\kappa_{\mathrm{transp}}=3L/8r$, which for our geometry leads to $\kappa_{\mathrm{transp}}=29.4$ \cite{Beijerink}. 

\section{Results and Analysis}

The properties of the source were measured using the setup illustrated in figure \ref{oven_layout}. Images of the fluorescence from the laser-illuminated atomic beam were taken using a CCD camera (\textit{Allied vision Guppy F038 B}) at a range of different laser detunings. In the case of caesium these are expressed relative to the $^{133}$Cs 6S$_{1/2}$ (F=4) $\rightarrow$ 6P$_{3/2}$ (F=5) transition (saturation intensity $I_{sat}= 2.71$ mW/cm$^2$), and in the case of lithium relative to the $^6$Li 2S$_{1/2}$ (F=3/2) $\rightarrow$ 2P$_{3/2}$ (F=5/2) transition (representative saturation intensity $I_{sat}= 2.54$ mW/cm$^2$, in practice multiple transitions can be addressed with slight variations). The probe beam is orthogonal to the direction of the atomic beam with a beam waist of $\omega= 1.3$ mm and a power of 0.47 mW for Li and 0.16 mW for Cs, corresponding to $I_{\text{max}}/I_{sat}=13.9$ for Li and $I_{\text{max}}/I_{sat}=4.5$ for Cs. 

\begin{figure}
\begin{center}
 \includegraphics[width = 17 cm]{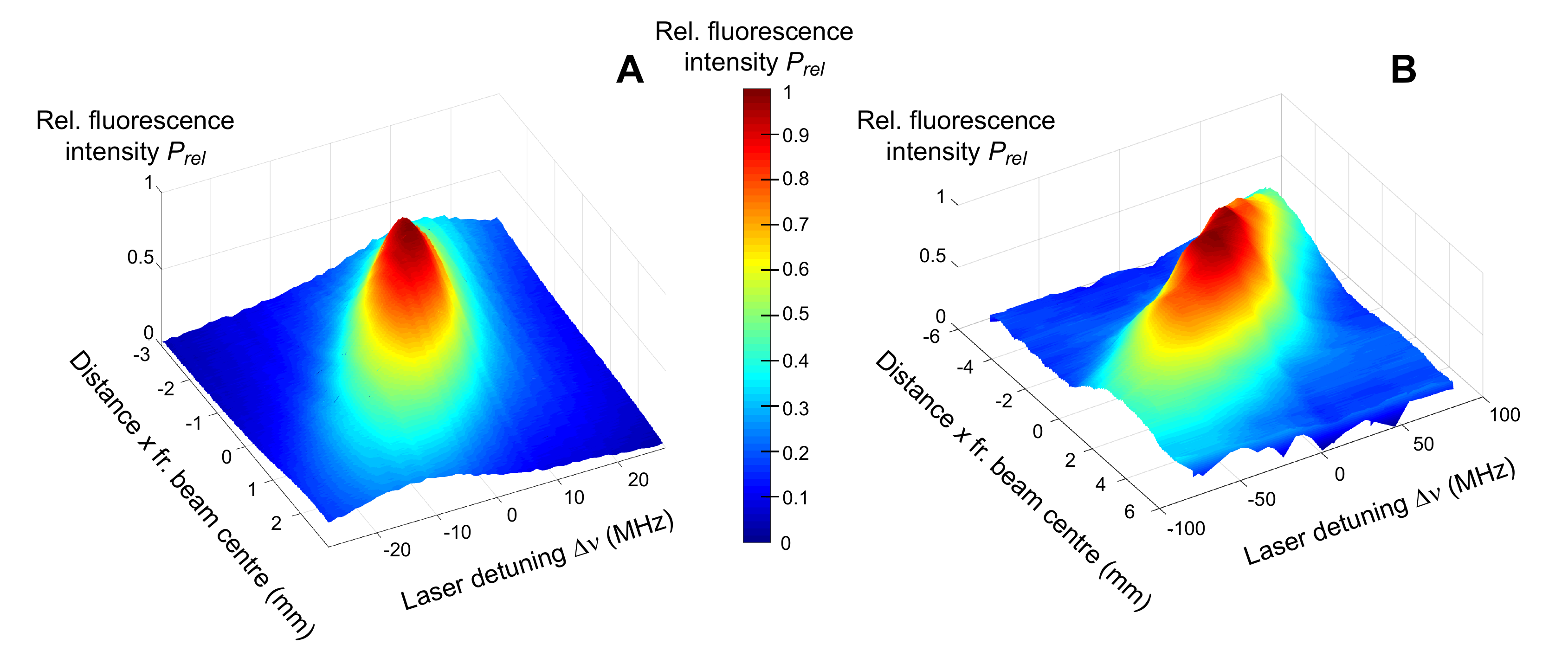}
 \caption{Surface plots showing the normalised fluorescence signal as a function of position $x$ and probe frequency detuning $\Delta \nu$ for caesium (A) and lithium (B).} 
 \label{surfs}
 \end{center}
\end{figure}

By taking images with spatial resolution using multiple probe laser detunings, two-dimensional data spanning the relevant frequency-position space were obtained. This contrasts with most previous beam characterisation studies, where only one-dimensional data is typically collected \cite{nonspat1,nonspat2,nonspat3}. In principle data of this form can be used to obtain the full angle-velocity distribution of the emitted particles, $I_{v,\theta}(v,\theta)$. The analysis we employ provides most of the key beam properties while being robust and practical in the presence of experimental noise. The raw data consists of a set of pixel values for each image, which are summed along the axis corresponding to the $z$ direction in figure \ref{oven_layout} to obtain a number of counts per pixel column as a function of both position along the $x$ direction and laser detuning $\Delta \nu$. The resulting signal $P(x,\Delta \nu)$ is plotted in figure \ref{surfs}. Errors were dominated by the systematic error arising from uncertainty in imaging magnification of $\pm 2 \%$. Uncertainty in frequency measurements is $\pm$0.1 MHz.

\begin{figure}
 \begin{center}
 \includegraphics[width = 10 cm]{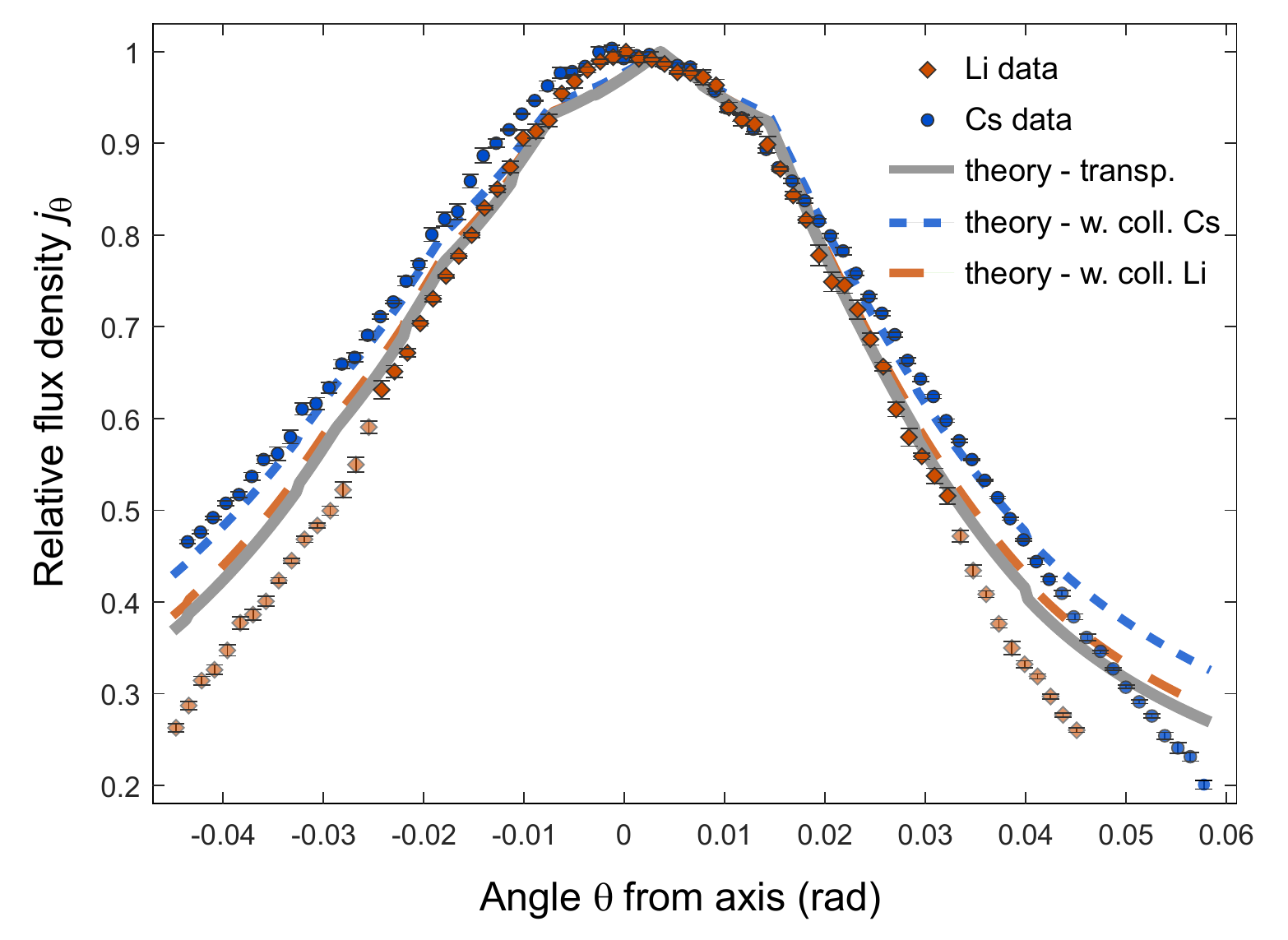}
 \caption{Angular distribution of the atoms emitted from the source. Points represent experimental data and the lines correspond to theoretical predictions for the relevant species using equations (\ref{Itheta}-\ref{endtheo}) with the broken lines showing the opaque regime, under the operating conditions described in \S 2. Blue represents Cs, orange Li and the grey line shows the theoretical prediction for the perfectly transparent case $K_{nL} = \infty$. The shaded points represent those where not all atoms were counted due to the limitation in frequency space (see text for details). Error bars represent statistical errors.}
 \label{plantainz}
 \end{center}
 
\end{figure}

To process this data, we first assume that the angle and speed of a particle's emission are uncorrelated, which has been shown to be a good approximation in similar systems \cite{scoles, olander, millerkush1}. This makes the total fluorescence counts recorded on a column of pixels, summed over all laser detunings, proportional to the flux density of the atoms at the corresponding $x$ position. From this, one obtains the relative flux density of atoms as a function of position in the observation plane, $j_x(x)$. In the limit of the source aperture being small compared to the distance to the observation plane, position in the observation plane can be directly converted to angle of emission. We also employ the standard approximation of Gaussian transverse velocity distributions \cite{microtube1}, together with the fact that the probe laser beam is narrower than the atomic beam, to allow us to treat $I_{x,y}(x,y)$ as separable in $x$ and $y$. This data therefore provides a good approximation to the angular distribution of the emitted particles, and an upper limit on the divergence of the particle beam. Figure \ref{plantainz} shows this data together with a theoretical prediction based on the analysis given in \S 2. The theoretical prediction was obtained by convolving the distribution $I_{\theta}(\theta)$ given by equations (\ref{Itheta}-\ref{endtheo}) with an array of delta functions corresponding to the positions of the microtubes within our nozzle, thus allowing us to avoid having to approximate the entire source aperture as small for this purpose. 

The vapour pressures of the two species were calculated based on their measured reservoir temperatures of $T_{\mathrm{Cs}}=133^{\circ}$C and $T_{\mathrm{Li}}=370^{\circ}$C, where the vapour pressure of Cs in the mixing chamber is reduced relative to that in the Cs reservoir by a factor of 0.641 as a result of the relative conductances of the output and mixing nozzles. This leads to an estimate of the mean free path lengths $\lambda_{P,Cs}=7.6$ mm and $\lambda_{P,Li}=29.5$ mm, i.e. for both species the source is expected to operate in the opaque regime, although the situation for lithium is expected to be close to the transparent regime with still $\lambda_{P,Li}>L$. From this, the relevant values of $K_{nL}$ for the theoretical predictions were determined.

Figure \ref{plantainz} shows a good general agreement between theory and experiment, with $R^2 = 0.98$ for Li and $R^2 = 0.99$ for Cs (when neglecting shaded data points --- see below). Lithium emission is still well described by the transparent case (grey line), but Cs shows clear, systematic deviations from the transparent case and only matches the theory when interparticle collisions are taken into account. For larger angles of emission the observed flux decreases faster than expected, even for the transparent case. This is an artefact resulting from the limited tuning range of our probe lasers leading to a truncation of the data in frequency space. Affected data points are shaded in the figure. 

An experimental value for the peaking factor can be derived by setting $\lambda_{P,a}$ and $\beta$ as free parameters and fitting (\ref{jthetalong}) to the experimental data. Substituting  their values into (\ref{kappa}) then yields peaking factors of $\kappa = 30 \pm $2 and $\kappa = 26 \pm $2 for Li and Cs respectively. These are in good agreement with the theoretical values of 29 and 24 given in \S2.

The total particle emission rates are estimated theoretically by integrating the expression for $I_{\theta}$ given in equations (\ref{Itheta}--\ref{endtheo}) over a half-sphere, using the measured reservoir temperatures and corresponding vapour pressures for the relevant species. This results in total emission rates of 1.2$\times 10^{13}$ atoms s$^{-1}$ for Li and 2.9$\times 10^{14}$ atoms s$^{-1}$ for Cs. Given the peaking factors this gives an estimated axial brightness of 3.6$\times 10^{14}$ atoms s$^{-1}$ sr$^{-1}$ for Li and 7.4$\times 10^{15}$ atoms s$^{-1}$ sr$^{-1}$ for Cs.

We can now also characterise the atomic velocity distribution within our beam by using our spectroscopic data. The fluorescence rate per atom depends on the detuning of the probe laser $\Delta \nu$ and the Doppler shift as a result of the transverse atomic velocity $v_x$. We consider an `effective detuning' ($\delta_{\text{eff}}=2\pi\Delta \nu - k v_x$) for each atom, with $k$ the wavevector of the probe laser, and the fluorescence rate per atom for a light intensity $I_{l}(y,z)$ is given by the standard expression \cite{foot}
\begin{equation}
 \Gamma_{sc,y,z}(\delta_{\text{eff}},y,z) = \frac{\Gamma}{2} \frac{I_l(y,z)/I_{\mathrm{sat}}}{\frac{4}{\Gamma^2}\delta_{\mathrm{eff}}^2+I_l(y,z)/I_{\mathrm{sat}}+1}
 \label{standard_yz}
\end{equation}
where $I_{\mathrm{sat}}$ is the saturation intensity and $\Gamma$ the natural linewidth of the relevant transition. In order to remove the $y$ and $z$ dependence of $\Gamma_{sc,y,z}(\delta_{\text{eff}},y,z)$, we assume that $v_x$ and $y$ are uncorrelated, which is consistent with existing theoretical models and results \cite{scoles, atmol_beams}. We also neglect the $z$ dependence of the atomic beam parameters since the probe laser beam is much smaller than the distance from the source to the observation plane. This allows us to average the scattering rate over $y$ and $z$:
\begin{equation}
 \Gamma_{sc}(\delta_{\text{eff}}) = \int_{-\infty}^{+\infty} \Gamma_{sc,y,z}(\delta_{\text{eff}},y,z) j_{y}(y) \,dy \,dz, \label{standard}
\end{equation}
where due to the cylindrical symmetry of the atomic beam $j_{y}(y)$ can be replaced with the already measured $j_x(x)$.

Approximating the transverse particle velocity distribution $f_{v_x}(v_x)$ as a Gaussian distribution \cite{microtube1}, the theoretical form for the total fluorescence $F(\Delta \nu)$ is the convolution of the fluorescence rate $\Gamma_{\text{sc}}(\delta_{\text{eff}})$ with the transverse velocity distribution $f_{v_x}(v_x)$:
\begin{equation}
 F(\Delta \nu) ~ \propto ~ \int_{-\infty}^{\infty} f_{v_x}(v_x) \Gamma_{\text{sc}}(\delta_{\text{eff}}) ~ d v_x.
 \label{trans_v}
\end{equation}
Experimental values for $F(\Delta \nu)$ from $P(x, \Delta \nu)$ can be obtained by summing the pixel counts from all $x$-positions in an image for a given frequency step. By fitting (\ref{trans_v}) to the measured $F(\Delta \nu)$, it is thus possible to obtain the Gaussian approximation of the transverse velocity distribution $f_{v_x}(v_x)$. 

\begin{figure}
 \begin{center}
 \includegraphics[width = 15 cm]{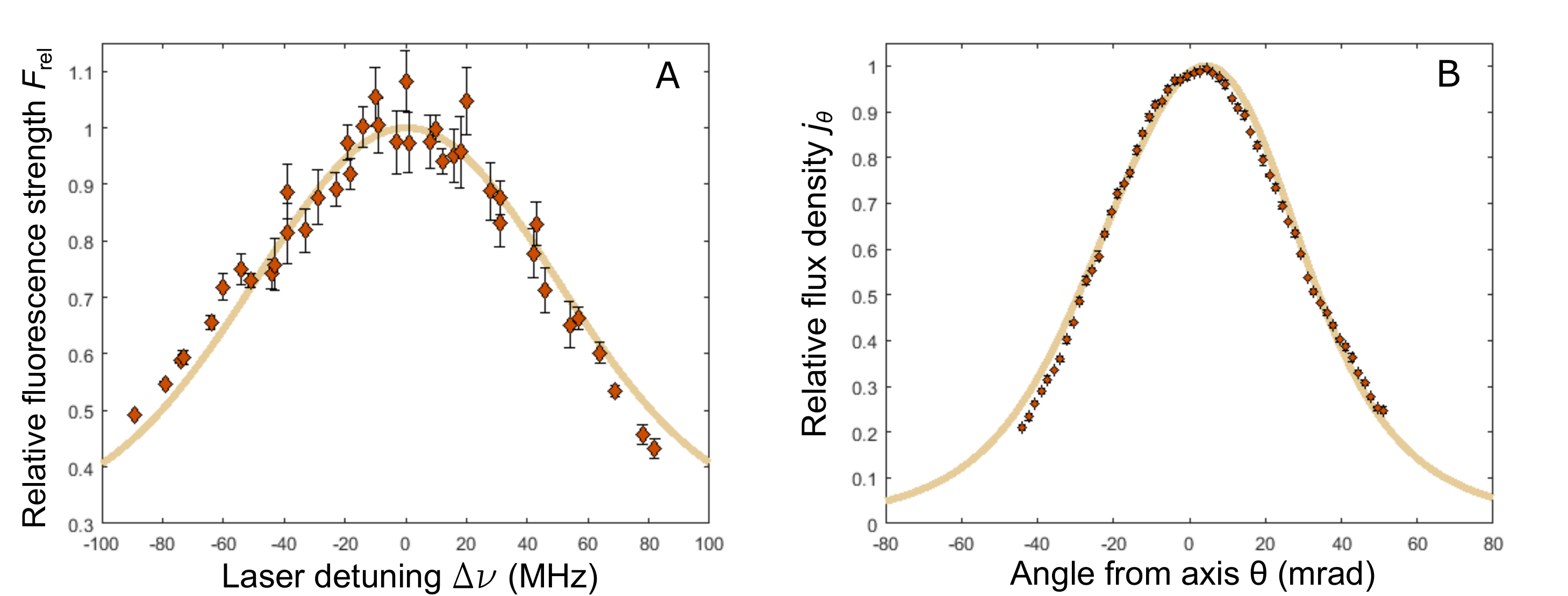}
 \caption{Frequency response of the Li fluorescence (A) and the angular distribution of the Li atomic flux (B). Points are experimental data and lines are a theoretical fit according to equations (\ref{trans_v}) and (\ref{long_conv}). Error bars represent statistical errors.}
 \label{lidbl}
 \end{center}
 
\end{figure}

\begin{figure}
 \begin{center}
 \includegraphics[width = 15 cm]{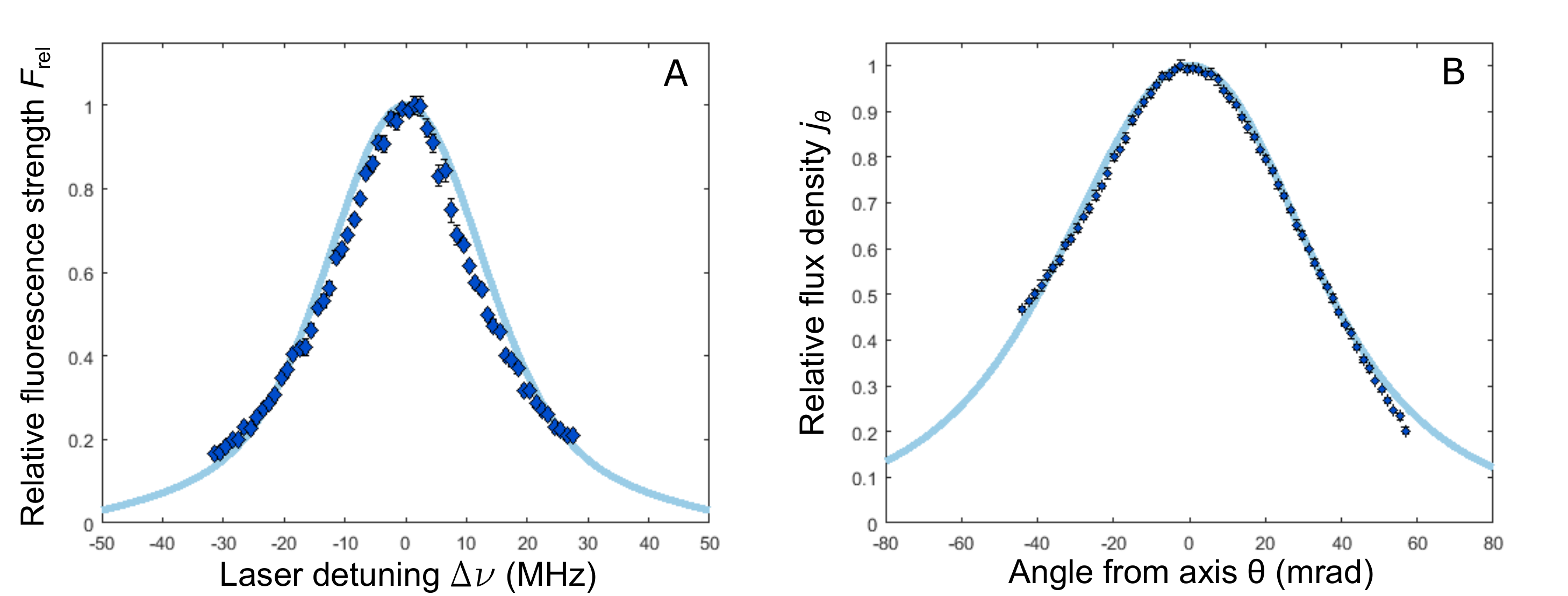}
 \caption{Frequency response of the Cs fluorescence (A) and the angular distribution of the Cs atomic flux (B). Points are experimental data and lines are a theoretical fit according to equations (\ref{trans_v}) and (\ref{long_conv}). Error bars represent statistical errors.}
 \label{csdbl}
 \end{center}
 
\end{figure}

Figures \ref{lidbl}(a) and \ref{csdbl}(a) show the measured fluorescence $F(\Delta \nu)$ and a theoretical fit according to (\ref{trans_v}) for Li and Cs respectively. A small systematic deviation can be seen between the Cs data and the fit in figure \ref{csdbl}(a), which shows that the measured transversal velocity distribution is close to, but not completely Gaussian.

In order to gain information about the longitudinal particle velocity distribution, $f_{v_z}(v_z)$, we start with a modified Gaussian distribution \cite{scoles,olander},  
\begin{equation}
f_{v_z}(v_z) = A v_z^3 \exp(-mv_z^2/(2k_B T_{z})).
\label{vz_dist}
\end{equation}
For a particle to emerge at an angle $\theta$ the ratio of its transverse to longitudinal velocities must be equal to $\tan{\theta}=v_x/v_z$. In order to obtain the angular output distribution $I_{\theta}(\theta)$ from $f_{v_x}$ and $f_{v_z}$ one can therefore use
\begin{equation}
 I_{\theta}(\theta) ~ \propto ~ \int_0 ^{\infty} f_{v_z}(v_z) f_{v_x}(v_z \tan{\theta}) v_z \sec{\theta} ~dv_z,
 \label{long_conv}
\end{equation}
where the $v_z \sec{\theta}$ term, representing a particle's total speed, arises from the variation of $d\theta/dv_{x}$, $d\theta/dv_{z}$ and the corresponding relations between density per unit angle and density per unit speed. It is now possible to determine the values of $A$ and $T_z$ from a fit of (\ref{long_conv}) to the measured angular distribution data, where the transverse velocity distribution $f_{v_x}$ was already obtained above by fitting (\ref{trans_v}) to the measured frequency response of the atomic fluorescence. The results of this fitting are plotted in figures \ref{lidbl}(b) and \ref{csdbl}(b). The resulting longitudinal temperatures are $T_{z}=(533\pm30)$ K for lithium and $T_{z}=(609\pm30)$ K for caesium, in reasonable agreement with externally measured microtube nozzle temperatures of 597 K during the Li measurements and 586 K during the Cs measurements. The fitted $T_z$ for Lithium is slightly lower than expected, and the difference between the measured and fitted temperature values may have been contributed to by temperature gradients within the device and the use of ex-vacuum temperature sensors. The dominant source of error in $T_z$ was the magnification uncertainty of $\pm 2$\% in the imaging system, the effects of which greatly exceeded those of the shot-to-shot variations and frequency uncertainty that affected the fitting of the transverse velocity distribution. 

Combining the measured longitudinal velocity distribution $f_{v_z}$ with a measurement of the spatial density distribution of particles in the observation plane, one can obtain absolute values for the flux density $I_x(x)$ and the total particle emission rate. These were calculated from the measured spatially dependent fluorescence data $P(x,\Delta \nu)$ using the relation
\begin{eqnarray}
\fl P(x,\Delta \nu) = \eta \int_{0}^{+\infty}\int_{-\infty}^{+\infty} \frac{1}{v_z} I_{x,v_x,v_z}(x, v_x, v_z) \Gamma_{sc}(2 \pi \Delta \nu - k v_x) ~d v_x ~dv_z,
\label{general_fluor}
\end{eqnarray}
where $\eta$ is a constant given by the product of the efficiency of the collection optics with the sensitivity of the CCD camera pixels (mean number of counts out per photon). Using the assumption that a particle's angle and speed of emission are uncorrelated, the integration can be separated to give
\begin{eqnarray}
\fl P(x,\Delta \nu) = \eta \int_{0}^{+\infty} \frac{1}{v_z} f_{z}(v_z)~dv_z~ \int_{-\infty}^{+\infty} I_{x,v_x}(x, v_x) \Gamma_{sc}(2 \pi \Delta \nu - kv_x) ~d v_x.
\label{simple_fluor}
\end{eqnarray}
Integrating (\ref{simple_fluor}) over laser detuning removes the $v_x$ dependence of the integral of $\Gamma_{sc}$, as $v_x$ only enters via a term of the form $(l = 2 \pi \Delta \nu - k v_x)$:
\begin{eqnarray}
 \fl \int_{-\infty}^{\infty} P(x,\Delta \nu)\,d\Delta \nu &= \eta \int_{0}^{+\infty} \frac{1}{v_z} f_{z}(v_z)\,dv_z \int_{-\infty}^{+\infty}I_{x,v_x}(x,v_x)dv_{x}\,\int_{-\infty}^{+\infty} \Gamma_{sc}(l) ~dl \\
 &= \eta I_{x}(x) \int_{0}^{+\infty} \frac{1}{v_z} f_{z}(v_z)\,dv_z \,\int_{-\infty}^{+\infty} \Gamma_{sc}(l) ~dl \\
 &= \eta I_{x}(x) \sqrt{\frac{\pi m}{8k_BT_z}}\int_{-\infty}^{+\infty} \Gamma_{sc}(l) ~dl
 \label{freqsum}
\end{eqnarray}
where we have replaced the integral of $I_{x,v_x}(x,v_x)$ over $v_x$ by $I_x(x)$, as these are equal by definition. Rearranging this yields 
\begin{equation}
 I_x(x) = \zeta \left( \eta \sqrt{\frac{\pi m}{8k_BT_z}} \int_{-\infty}^{+\infty} \Gamma_{sc}(l)~dl  \right)^{-1} 
 \Upsilon \sum_i P(x,\Delta \nu_i),
\end{equation} 
where $\Upsilon$ is the difference in laser detuning between successive images and the remaining integral over $\Gamma_{sc}(l)~dl$ can be evaluated numerically. To get $I_{\theta}$ from $I_x$ one can simply use the relation $\tan{\theta} = x/D$, with $D$ the distance from the source to the observation plane.  

The factor $\zeta$ is introduced post-hoc to allow for attenuation of the fluorescence light by a metallic coating which formed on the inside of the viewport used to provide optical access for the camera. While direct measurement of $\zeta$ was not possible without major experimental modifications, it is still possible to check for consistency between the results for Li and Cs. For Cs the expected total emission rate is $2.85 \times 10^{14}$ s$^{-1}$, while the measured rate with $\zeta = 1$ is $2.2 \times 10^{12}$ atoms s$^{-1}$, implying a value of $\zeta$ of 130. For Li it is $1.2 \times 10^{13}$ s$^{-1}$, while the measured rate with $\zeta = 1$ is $8.6 \times 10^{10}$ atoms s$^{-1}$, giving a value of $\zeta$ of 140. These values are consistent to within experimental error. They therefore indicate that the ratio of the total Li flux to the total Cs flux matches that expected given the theory described in \S2.   

We expect that this problem could be avoided in future designs by providing better thermal shielding for the camera, thus allowing the viewport to be heated to a higher temperature, or by using UV light to clear the viewport \cite{liad}. Alternatively the viewports could be moved further from the atomic beam, or a cold plate could be used (as in \cite{bigbec}, for example) to encourage condensation of stray atoms in a specific area and thus avoid coating of room temperature surfaces. 

\section{Summary}

We have demonstrated a dual-species oven based on a microtube array and shown that it is an effective device for sourcing fully overlapping, collimated beams of Li and Cs. The measured peaking factors of $\kappa=30 \pm 2$ and $\kappa=26 \pm 2$ for Li and Cs respectively are in good agreement with theoretical values. The result is an improvement on the degree of collimation actually achieved in some previous experiments, where imperfect alignment of the microtubes was found to increase the angular spread of the output \cite{microtube1}. We have successfully used the existing single species theory for a multi-species situation and found good agreement. 

The spatially resolved fluorescence spectroscopy that we used to characterise the device allows us to experimentally determine both longitudinal and transverse beam properties, and combined with the analysis given in \S3 it provides an experimentally convenient way to determine the key properties of an atomic beam. Furthermore, the method does in principle allow the measurement of more complex beam properties such as correlations between the speed and angle of particle emission. This could be particularly interesting for the study of sources operating in a higher pressure regime, where collisional and fluid dynamical effects may occur. It should also be possible to use the data collected with spatially resolved fluorescence spectroscopy to recover full particle velocity distributions, without having to make any prior assumptions about their mathematical form. In order to do this, a larger data set would be required so that deconvolution methods could be used effectively despite the experimental noise, and carrying out such a study is a potential future direction for this work. 

Another possible refinement of the current work would be the use of a wider range of laser detunings, particularly when studying the emission of Lithium, in order to avoid frequency-truncation artefacts at higher angles of emission.

\section{Acknowledgments}

The authors would like to thank Mike Robson for his contribution to the image analysis code used in this work. 
This project was supported by the  EPSRC grants EP/K023624/1, EP/M013294/I and by the European Comission grant QuILMI - Quantum Integrated Light Matter Interface (No 295293). We acknowledge support from the University of Nottingham through a Birmingham-Nottingham collaboration grant.

\section{References}


\bibliographystyle{unsrt}
\bibliography{ovenbib_updated_fix}

\appendix

\section{Technical construction details}

Heating was carried out using Isomil mineral insulated heating wire, with separate wires used for each of the mixing chamber, microtube nozzle, Cs reservoir and Li reservoir. The currents in the wires for the mixing chamber and microtube nozzle were set manually, while in the case of the Cs and Li reservoirs thermistors were introduced, well thermally contacted to the outside of the vacuum apparatus, to allow readout and feedback control of the reservoir temperatures. Feedback control was performed using a PI loop implemented via an Arduino Uno microcontroller board, which controlled the gate pin voltage of a pair of TIP120G silicon transistors connected in series with each of the two reservoir heating wires.

In order to ensure good alignment of the microtubes 15 tubes were mounted inside an equilateral triangle of appropriate dimension. To provide this support structure 3 component parts were made and bolted together such that they formed a cylinder of outer diameter equal to the inner diameter of the vacuum tube, with a void of triangular cross section running along the centre. By combining three separate pieces it was possible to achieve sharp corners on the triangle, which it would otherwise have been difficult to machine at the relevant length scale, and tightening of the connecting bolts ensured that the microtubes were well secured. To avoid the need for vacuum screws, only through-holes were used to bolt the parts together.



\end{document}